\documentclass[journal,comsoc]{IEEEtran}
\usepackage[T1]{fontenc}
\usepackage{color}
\usepackage{amsmath}
\usepackage{array}
\usepackage{fancyhdr}
\usepackage{latexsym}
\usepackage{epsfig}
\usepackage{amssymb}
\usepackage{amsfonts}
\usepackage{amsxtra}
\usepackage{xspace}
\usepackage{makeidx}
\usepackage{graphics}
\usepackage{theorem}
\usepackage{subfigure}
\usepackage{algorithm,algpseudocode}
\usepackage{multirow}

\newtheorem{proposition}{\textbf{Proposition}}

\begin{document}

\title{Hybrid MMSE Beamforming for Multiuser Millimeter Wave Communication Systems}

\author{\IEEEauthorblockN{Jiaqi Cong, Tian Lin, and Yu Zhu,~\IEEEmembership{Member,~IEEE}}

\thanks{Manuscript received May 30, 2018; revised August 2, 2018; accepted August 27, 2018. This work was supported in part by National Natural Science Foundation of China under Grant No. 61771147. The associate editor coordinating the review of this paper and approving it for publication was D. Cassioli. \textit{(Corresponding author: Yu Zhu.)}}


\thanks{J. Cong, T. Lin, and Y. Zhu are with the State Key Laboratory of ASIC and System, Department of Communication Science and Engineering, Fudan University, Shanghai, China (e-mail: jqcong16@fudan.edu.cn; lint17@fudan.edu.cn; zhuyu@fudan.edu.cn).}}

\maketitle

\begin{abstract}
This letter investigates the hybrid analog and digital beamforming (HBF) design for multiuser millimeter wave (mmWave) communication systems based on the minimum mean square error (MMSE) criterion. Using the alternating minimization method, the hybrid precoder of the base station (BS) and the hybrid combiners of the users are alternatively optimized. It is shown that both the optimized digital precoder of the BS and the digital combiners of the users have closed-form expressions, and their corresponding analog ones can be efficiently obtained via generalized eigen-decomposition. Simulation results show that the proposed MMSE HBF scheme has fast convergence and performs close to the fully digital beamforming.
\end{abstract}


\IEEEpeerreviewmaketitle

\section{Introduction}
Hybrid analog and digital beamforming (HBF) design has recently been recognized as a key technology in millimeter wave (mmWave) communication systems to improve the spectral efficiency and/or energy-efficiency at affordable hardware cost and power consumption \cite{sohrabi2016hybrid}-\cite{Tsinos2017on}. Although its application to multiuser multiple input and multiple output (MIMO) mmWave systems enables spatial division multiple access, there also exist big challenges since the signals at different users cannot be cooperatively processed \cite{sohrabi2016hybrid},\cite{Payami2016hybrid}-\cite{cong2017hybrid}.


In the existing studies on the multiuser HBF design, the authors in \cite{sohrabi2016hybrid} and \cite{Payami2016hybrid} investigated the HBF design in the multiuser multiple input and single output (MISO) scenario aiming at maximizing the sum achievable rate. In \cite{alkhateeb2015limited}, the authors proposed a low-complexity HBF scheme in the multiuser MIMO scenario supporting multiple data streams for each user. More recently, the authors in \cite{nguyen2017hybrid} investigated the the orthogonal matching pursuit (OMP) based HBF algorithm under the minimum mean square error (MMSE) criterion. To enhance the performance, in \cite{cong2017hybrid}, the authors proposed a near-optimal multiuser MMSE HBF scheme in MISO scenario.

In this paper, we investigate the HBF design aiming to minimize the sum of the mean square errors (sum-MSE) of all users' multiple streams in a downlink multiuser MIMO mmWave system \footnote{As shown in the traditional fully digital MIMO beamforming designs \cite{palomar2003joint}, the objective of minimizing the sum-MSE results in fairer beamforming and power allocation among data streams than that of maximizing the sum-rate.}. Using the alternating minimization method \cite{Csi1984Alt}, we decompose the original problem into the hybrid precoding and combining sub-problems. For the former sub-problem, we derive the optimal digital precoder based on the  Karush-Kuhn-Tucker (KKT) conditions and optimize the analog one via generalized eigen-decomposition (GEVD). For the latter one, we derive a closed-form expression of the digital combiners under the unitary constraint and optimize the analog combiners via GEVD by replacing the sum-MSE by its lower bound. Simulation results show that the proposed MMSE HBF scheme outperforms the conventional HBF schemes and performs close to the fully digital beamforming.

\textit{Notations}: $\mathbf{A}$ is a matrix, $\mathbf{a}$ is a vector, and $a$ is a scalar. $\mathbf{I}_N$ is an $N\times N$ identity matrix. $\mathrm{blkdiag}\lbrace \mathbf{A}_1, \mathbf{A}_2, \dots, \mathbf{A}_N\rbrace$ returns a block diagonal matrix with sub-matrices $\mathbf{A}_1, \mathbf{A}_2, \dots, \mathbf{A}_N$ on its diagonal. $\mathbf{A}^T$, $\mathbf{A}^H$ and $\mathbf{A}^{-1}$ are the transpose, conjugate transpose and inverse of matrix $\mathbf{A}$. $\mathrm{tr}\left(\mathbf{A} \right)$ denotes the trace of matrix $\mathbf{A}$. $\mathrm{Re}\lbrace\cdot\rbrace$ denotes the real component of a complex variable. $\|\cdot\|_1$, $\|\cdot\|$ and $\|\cdot\|_\infty$ are the one, two and infinite norms, respectively. $\mathcal{CN}\left(\mathbf{a},\mathbf{A}\right)$ denotes the circularly symmetric complex Gaussian distribution with mean $\mathbf{a}$ and covariance matrix $\mathbf{A}$. $\mathrm{E}\{\cdot \}$ denotes the expectation operator.

\section{System Model}\label{sec:systemmodel}
\begin{figure}[t]
\begin{center}
\centering
\includegraphics*[width=3.5in]{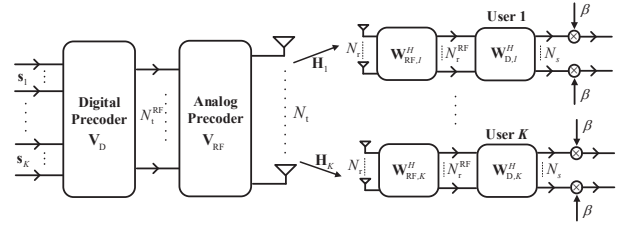}
\caption{Diagram of the downlink of a multiuser MIMO mmWave system with the hybrid precoding and combining architecture.}
\label{model}
\end{center}
\end{figure}
Consider the downlink of a narrowband multiuser mmWave MIMO system shown in Fig. \ref{model}, where a base-station (BS) with $N_\mathrm{t}$ transmit antennas and $N_\mathrm{t}^\mathrm{RF}$ RF chains serves a total of $K$ users each of which is equipped with $N_\mathrm{r}$ receive antennas and $N_\mathrm{r}^\mathrm{RF}$ RF chains and requires $N_\mathrm{s}$ independent data streams. It is assumed that $KN_\mathrm{s}\leq N_\mathrm{t}^\mathrm{RF}\ll N_\mathrm{t}$ and $N_\mathrm{s}\leq N_\mathrm{r}^\mathrm{RF}\leq N_\mathrm{r}$ due to the high cost and power consumption of RF devices. Throughout this letter the fully connected RF precoder/combiner structure \cite{yu2016alternating} is considered. At the BS, the users' data streams are processed with a baseband precoder $\mathbf{V}_\mathrm{D}$ followed by an RF preocoder $\mathbf{V}_\mathrm{RF}$. Thus, the precoded signal is given by
$\mathbf{x} = \mathbf{V}\mathbf{s} = \mathbf{V}_\mathrm{RF}\mathbf{V}_\mathrm{D}\mathbf{s} = \sum_{k=1}^K\mathbf{V}_\mathrm{RF}\mathbf{V}_{\mathrm{D},k}\mathbf{s}_{k}$,
where $\mathbf{V}=\mathbf{V}_\mathrm{RF}\mathbf{V}_\mathrm{D}$ denotes the hybrid precoding matrix with $\mathbf{V}_\mathrm{D}=\left[\mathbf{V}_\mathrm{D,1},\dots,\mathbf{V}_{\mathrm{D},K}\right]$ and $\mathbf{V}_{\mathrm{D},k}$ being an $N_\mathrm{t}^\mathrm{RF}\times N_\mathrm{s}$ matrix for $k=1,\dots,K$, and $\mathbf{s}=[\mathbf{s}_{1}^T,\dots,\mathbf{s}_{K}^T]^T$ with $\mathrm{E}\{\mathbf{s}\mathbf{s}^H\}=\mathbf{I}_{KN_\mathrm{s}}$ is the $KN_\mathrm{s}\times1$ vector of all users' transmitted symbols, with $\mathbf{s}_{k}$ defined as the symbol vector of user $k$. Furthermore, it is assumed that $\mathrm{tr}\left(\mathbf{V}_\mathrm{RF}\mathbf{V}_\mathrm{D}\mathbf{V}_\mathrm{D}^H\mathbf{V}_\mathrm{RF}^H\right)\leq P$, where $P$ is the maximum transmit power of the BS. 

Assuming a frequency-flat fading MIMO channel between the BS and user $k$, the received signal vector at user $k$ is $\mathbf{y}_{k} = \mathbf{H}_k\mathbf{x}+\mathbf{e}_{k}$,
where $\mathbf{e}_{k}\sim\mathcal{CN}\left(0,\sigma^2\mathbf{I}_{N_\mathrm{r}}\right)$ denotes the noise vector, and the channel response $\mathbf{H}_k$ is modeled as
\begin{equation}
\mathbf{H}_k=\sqrt{\frac{N_\mathrm{t}N_\mathrm{r}}{L}}\sum\limits_{l=1}^L\alpha_{l,k}\mathbf{a}_\mathrm{r} \left(\phi_{\mathrm{r},k}^l\right)\mathbf{a}_\mathrm{t}^H\left(\phi_{\mathrm{t},k}^l\right),
\label{channel}
\end{equation}
where $\alpha_{l,k}$, $\phi_{\mathrm{r},k}^l$ and $\phi_{\mathrm{t},k}^l$ denote the complex gain, the angles of departure and arrival (AoD and AoA) corresponding to the $l$th path, respectively. Further, $\mathbf{a}_\mathrm{r}\left(.\right)$ and $\mathbf{a}_\mathrm{t}\left(.\right)$ are the antenna array response vectors at the BS and a user, respectively. Considering the uniform linear arrays, we have
$\mathbf{a}_i\left(\phi\right)=\frac{1}{\sqrt{N_i}}\left[1,e^{\mathrm{j}k_0d\sin\left(\phi\right)},\dots, e^{\mathrm{j}k_0d\left(N_i-1\right)\sin\left(\phi\right)}\right]^T$,
where $i\in \{ \mathrm{r},\mathrm{t}\}$, $\mathrm{j}=\sqrt{-1}$, $k_0=2\pi/\lambda_c$, $\lambda_c$ is the wavelength, and $d$ is the antenna spacing. It is assumed that $\mathbf{H}_1,\dots,\mathbf{H}_K$ are perfectly known at the BS.

For each user, the received signal is first processed with an analog combiner $\mathbf{W}_{\mathrm{RF},k}$, then a low-dimensional digital combiner $\mathbf{W}_{\mathrm{D},k}$, and finally a symbol estimator denoted by a scalar factor $\beta$ \cite{joung2007regularized}. That is,
\begin{equation}
\begin{split}
\mathbf{\hat{s}}_{k}=&\beta\mathbf{W}_{\mathrm{D},k}^H\mathbf{W}_{\mathrm{RF},k}^H\mathbf{H}_k\mathbf{V}_k \mathbf{s}_{k}+\beta\mathbf{W}_{\mathrm{D},k}^H\mathbf{W}_{\mathrm{RF},k}^H\mathbf{H}_k\sum_{f\neq k}\mathbf{V}_f\mathbf{s}_{f}\\
&+\beta\mathbf{W}_{\mathrm{D},k}^H\mathbf{W}_{\mathrm{RF},k}^H\mathbf{e}_{k},\nonumber
\end{split}
\end{equation}
where the three terms in the right hand side represent the desired signal, the inter-user interference and the noise, respectively. Define the MSE of user $k$ as $J_k=\mathrm{E}\{||\mathbf{s}_{k}-\mathbf{\hat{s}}_{k}||^2\}$. By substituting the above equation into this definition, we have
\begin{equation}
\begin{split}
J_k=&\mathrm{tr}\left(\beta^2\mathbf{W}_k^H\mathbf{H}_k\mathbf{V}\mathbf{V}^H\mathbf{H}_k^H\mathbf{W}_k+ \beta^2\sigma^2\mathbf{W}_k^H\mathbf{W}_k+\mathbf{I}_{N_\mathrm{s}}\right)\\ &-2\mathrm{Re}\lbrace\mathrm{tr}\left(\beta\mathbf{V}_k^H\mathbf{H}_k^H\mathbf{W}_k\right)\rbrace,
\end{split}
\label{sumMSE}
\end{equation}
where $\mathbf{W}_k=\mathbf{W}_{\mathrm{RF},k}\mathbf{W}_{\mathrm{D},k}$ and $\mathbf{V}_k=\mathbf{V}_\mathrm{RF}\mathbf{V}_{\mathrm{D},k}$. Since $\mathbf{V}_\mathrm{RF}$ and $\mathbf{W}_{\mathrm{RF},k}$ are implemented using phase shifters, we introduce the constant modulus constraint on each entry of the analog beamformers. The objective in this letter is to minimize the sum-MSE of all users' multiple streams. Thus, the HBF optimization problem is formulated as follows:
\begin{equation}
\begin{split}
\underset{\mathbf{V}_\mathrm{D},\mathbf{V}_\mathrm{RF},\mathbf{W}_{\mathrm{D},k},\mathbf{W}_{\mathrm{RF},k},\beta}{\text{minimize}}  & J_{\mathrm{sum}}=\sum_{k=1}^KJ_k \\
\text{subject to}\;\;\;\;\;\;\;\;\; & \mathrm{tr}\left(\mathbf{V}_\mathrm{RF}\mathbf{V}_\mathrm{D}\mathbf{V}_\mathrm{D}^H\mathbf{V}_\mathrm{RF}^H\right)\leq P\\
& |\mathbf{V}_\mathrm{RF}(\mathit{i},\mathit{j})|^2=1,\;\forall \mathit{i},\mathit{j} \\
& |\mathbf{W}_{\mathrm{RF},k}(\mathit{p},\mathit{q})|^2=1,\;\forall \mathit{p},\mathit{q},\mathit{k}.
\end{split}
\label{optproblem}
\end{equation}

\section{Hybrid MMSE Precoder and Combiners Design}\label{design}
As the problem in \eqref{optproblem} is nonconvex and difficult to solve optimally, based on the alternating minimization method, we propose a HBF scheme to alternatively optimize the hybrid precoder of the BS and the hybrid combiners of the users.

\subsection{Hybrid Precoder Design}\label{subsec:hybrid-prec}
By fixing all users' hybrid combiners, we have the following BS hybrid precoding optimization sub-problem:
\begin{equation}
\begin{split}
\underset{\mathbf{V}_\mathrm{D},\mathbf{V}_\mathrm{RF},\beta}{\text{minimize}}\;\;\;\;  & J_{\mathrm{sum}} \\
\text{subject to}\;\;\;\; & \mathrm{tr}\left(\mathbf{V}_\mathrm{RF}\mathbf{V}_\mathrm{D}\mathbf{V}_\mathrm{D}^H\mathbf{V}_\mathrm{RF}^H\right)\leq P\\
&|\mathbf{V}_\mathrm{RF}(\mathit{i},\mathit{j})|^2=1,\;\forall \mathit{i},\mathit{j},
\end{split}
\end{equation}
where the scalar factor $\beta$ is jointly optimized with $\mathbf{V}_\mathrm{D}$ and $\mathbf{V}_\mathrm{RF}$ for better performance since now the noise effect is considered in the precoder design.

\subsubsection{Digital Precoder Design}
We first fix $\mathbf{V}_\mathrm{RF}$ and optimize $\beta$ and $\mathbf{V}_\mathrm{D}$. As shown in \cite{cong2017hybrid}, the original precoder $\mathbf{V}_\mathrm{D}$ can be separated as $\mathbf{V}_\mathrm{D} =\beta^{-1}\widetilde{\mathbf{V}}_\mathrm{D}$, where $\widetilde{\mathbf{V}}_\mathrm{D}$ is an unconstrained baseband precoder and $\beta$ has the function of guaranteeing the transmit power constraint. Based on the KKT conditions, it can be shown that the optimal $\mathbf{V}_\mathrm{D}$ and $\beta$ are given by
\begin{equation}
\begin{split}
\widetilde{\mathbf{V}}_\mathrm{D}&=\left(\mathbf{V}_\mathrm{RF}^H\mathbf{H}^H\mathbf{W}\mathbf{W}^H\mathbf{H} \mathbf{V}_\mathrm{RF}+\lambda\mathbf{V}_\mathrm{RF}^H\mathbf{V}_\mathrm{RF}\right)^{-1}\mathbf{V}_\mathrm{RF}^H \mathbf{H}^H\mathbf{W},\\
\beta&=\sqrt{\mathrm{tr}(\mathbf{V}_\mathrm{RF}\widetilde{\mathbf{V}}_\mathrm{D} \widetilde{\mathbf{V}}_\mathrm{D}^H\mathbf{V}_\mathrm{RF}^H)/P},\nonumber
\end{split}
\end{equation}
where $\mathbf{H}=\left[\mathbf{H}_1^T, \dots, \mathbf{H}_K^T\right]^T$, $\lambda=\sigma^2\mathrm{tr}\left(\mathbf{W}^H\mathbf{W}\right)\big/P$, and $\mathbf{W}=\mathrm{blkdiag}\lbrace \mathbf{W}_1, \dots, \mathbf{W}_K\rbrace$ is a block diagonal matrix with all users' hybrid combining matrices on the diagonal.

\subsubsection{Analog Precoder Design}
By substituting the above optimal digital precoder into the sum-MSE and using the matrix inversion lemma, we have
\begin{equation}
\begin{split}
J_{\mathrm{sum}}=&\mathrm{tr}\Big(\big(\mathbf{I}_{KN_\mathrm{s}} +\frac{1}{\lambda} \mathbf{W}^H\mathbf{H}\mathbf{V}_\mathrm{RF}(\mathbf{V}_\mathrm{RF}^H\mathbf{V}_\mathrm{RF})^{-1}\\ &\quad\quad\times\mathbf{V}_\mathrm{RF}^H\mathbf{H}^H\mathbf{W}\big)^{-1}\Big),
\end{split}
\label{MSEequation}
\end{equation}
which is now a function of $\mathbf{V}_\mathrm{RF}$ to be further optimized. Due to the fact that the BS is equipped with a large number of transmit antennas, the analog beamforming vectors are likely orthogonal to each other \cite{sohrabi2016hybrid}, i.e., $\mathbf{V}_\mathrm{RF}^H\mathbf{V}_\mathrm{RF}\approx N_\mathrm{t}\mathbf{I}_{N_\mathrm{t}^\mathrm{RF}}$. Under this approximation and further using the Sherman Morrison formula, the sum-MSE in \eqref{MSEequation} can be separated into two terms that are related respectively to a column in $\mathbf{V}_\mathrm{RF}$, denoted by $\mathbf{v}^{(j)}_\mathrm{RF}$, and the remaining sub-matrix, denoted by $\overline{\mathbf{V}}^{(j)}_\mathrm{RF}$, after removing  $\mathbf{v}^{(j)}_\mathrm{RF}$ from $\mathbf{V}_\mathrm{RF}$. That is,
\begin{equation}
\begin{split}
J_{\mathrm{sum}}{\approx}&\mathrm{tr}\left(\left(\mathbf{I}_{KN_\mathrm{s}}+\frac{1}{\eta}\mathbf{W}^H \mathbf{H}\mathbf{V}_\mathrm{RF}\mathbf{V}_\mathrm{RF}^H\mathbf{H}^H\mathbf{W}\right)^{-1}\right)\\
=&\mathrm{tr}\left(\mathbf{A}_{\mathrm{t},j}^{-1}\right)-\frac{\mathbf{v}_\mathrm{RF}^{(j) H}\left(\frac{1}{\eta}\mathbf{H}^H\mathbf{W}\mathbf{A}_{\mathrm{t},j}^{-2}\mathbf{W}^H\mathbf{H}\right) \mathbf{v}_\mathrm{RF}^{(j)}}{\mathbf{v}_\mathrm{RF}^{(j) H}\left(\frac{1}{N_\mathrm{t}}\mathbf{I}+\frac{1}{\eta}\mathbf{H}^H\mathbf{W}\mathbf{A}_{\mathrm{t},j}^{-1} \mathbf{W}^H\mathbf{H}\right)\mathbf{v}_\mathrm{RF}^{(j)}},
\end{split}
\label{process_t}
\end{equation}
where $\mathbf{A}_{\mathrm{t},j}=\mathbf{I}+\frac{1}{\eta}\mathbf{W}^H\mathbf{H}\overline{\mathbf{V}}_\mathrm{RF}^{(j)} (\overline{\mathbf{V}}_\mathrm{RF}^{(j)})^H\mathbf{H}^H\mathbf{W}$  and $\eta=N_\mathrm{t}\lambda$. A close observation to \eqref{process_t} reveals that $\mathbf{V}_\mathrm{RF}$ can be optimized column-by-column. Specifically, $\mathbf{v}_\mathrm{RF}^{(j)}$ can be optimized by maximizing the last term in \eqref{process_t}. Define $\mathbf{B}_{\mathrm{t},j}=\frac{1}{\eta}\mathbf{H}^H\mathbf{W}\mathbf{A}_{\mathrm{t},j}^{-2}\mathbf{W}^H\mathbf{H}$ and $\mathbf{D}_{\mathrm{t},j}=\frac{1}{N_\mathrm{t}}\mathbf{I}+\frac{1}{\eta}\mathbf{H}^H \mathbf{W}\mathbf{A}_{\mathrm{t},j}^{-1}\mathbf{W}^H\mathbf{H}$. It can be shown that by fixing other columns of the RF precoder and ignoring the constant modulus constraint, the optimal $\mathbf{v}_\mathrm{RF}^{(j)}$ is the eigenvector associated with the largest generalized eigenvalue of the matrix pair $\mathbf{B}_{\mathrm{t},j}$ and $\mathbf{D}_{\mathrm{t},j}$. Considering the constant modulus constraint, a sub-optimal solution of $\mathbf{v}_\mathrm{RF}^{(j)}$ can be obtained by directly extracting the phase of each element of the eigenvector as similar to that in \cite{yu2016alternating,Payami2016hybrid}. Here the phase extraction is performed before the optimization of the next column, i.e., $\mathbf{v}_\mathrm{RF}^{(j+1)}$. Note that although the iteration convergence cannot be proved due to the phase extraction, simulation results in Section IV will show that the overall performance of the proposed HBF scheme converges fast.

\subsection{Hybrid Combiners Design}\label{subsec:hybrid-comb}
We now consider the hybrid combiners design with the optimized precoder. We first optimize the users' digital combiners by fixing the analog ones. Inspired by \cite{sohrabi2016hybrid,yu2016alternating}, a similar constraint that the columns of the digital combiner of user $k$ are mutually orthogonal is imposed. That is,
\begin{equation}
\mathbf{W}_{\mathrm{D},k}^H\mathbf{W}_{\mathrm{D},k}=\gamma\mathbf{I}_{N_\mathrm{s}},
\label{supposing}
\end{equation}
where $\gamma>0$. Note that $\gamma$ can be absorbed in the $\beta$ factor. Thus, in the following, $\gamma$ is set to 1 without loss of generality.

\subsubsection{Digital Combiners Design}
From the constraint \eqref{supposing}, it can be shown that $\mathbf{W}_{\mathrm{D},k}\mathbf{W}_{\mathrm{D},k}^H=\mathbf{Z}_k\begin{bmatrix} \mathbf{I}_{N_\mathrm{s}}&\mathbf{0}\\ \mathbf{0}&\mathbf{0} \end{bmatrix}\mathbf{Z}_k^H$, where $\mathbf{Z}_k$ is an $N_\mathrm{r}^\mathrm{RF}\times N_\mathrm{r}^\mathrm{RF}$ unitary matrix. By substituting this result into \eqref{sumMSE} and further fixing $\mathbf{V}_\mathrm{RF}$, $\mathbf{V}_\mathrm{D}$, $\mathbf{W}_{\mathrm{RF},k}$ and $\beta$ in \eqref{sumMSE}, it can be found that only the last term in \eqref{sumMSE} is a function of $\mathbf{W}_{\mathrm{D},k}$. By taking this observation into the objective function of \eqref{optproblem} and removing the terms that are not related to $\mathbf{W}_{\mathrm{D},k}$, the optimization problem \eqref{optproblem} is now converted into
\begin{equation}
\begin{split}
\underset{\mathbf{W}_{\mathrm{D},k}}{\text{maximize}} \;\;\; & \sum_{k=1}^K\mathrm{Re}\{\mathrm{tr}\left(\beta\mathbf{V}_{\mathrm{D},k}^H\mathbf{V}_\mathrm{RF}^H \mathbf{H}_k^H\mathbf{W}_{\mathrm{RF},k}\mathbf{W}_{\mathrm{D},k}\right)\} \\
\text{subject to}\;\;\; & \mathbf{W}_{\mathrm{D},k}^H\mathbf{W}_{\mathrm{D},k}=\mathbf{I}_{N_\mathrm{s}}, \text{for}\;k=1,\dots,K.
\end{split}\nonumber
\end{equation}
It turns out that this problem is still difficult to solve directly. Instead, the optimization can be carried out by aiming at its upper bound, which is $\sum_{k=1}^K\mathrm{Re}\{\mathrm{tr}(\mathbf{G}_k\mathbf{W}_{\mathrm{D},k})\}\leq \sum_{k=1}^K|\mathrm{tr}(\mathbf{G}_k\mathbf{W}_{\mathrm{D},k})|$, where $\mathbf{G}_k=\beta\mathbf{V}_{\mathrm{D},k}^H\mathbf{V}_\mathrm{RF}^H\mathbf{H}_k^H\mathbf{W}_{\mathrm{RF},k}$. By using the H$\ddot{\text{o}}$lder's inequality \cite{horn1990matrix}, we have
\begin{equation}
\sum_{k=1}^K\mid\mathrm{tr}\left(\mathbf{G}_k\mathbf{W}_{\mathrm{D},k}\right)\mid
\leq\sum_{k=1}^K\|\mathbf{W}_{\mathrm{D},k}^H\|_\infty\cdot\|\mathbf{G}_k\|_1.
\label{inequ}
\end{equation}
With the unitary constraint in \eqref{supposing}, we have $\|\mathbf{W}_{\mathrm{D},k}^H\|_\infty=1$. Taking the singular value decomposition (SVD) to $\mathbf{G}_k$, we have $\mathbf{G}_k=\mathbf{U}\Sigma\mathbf{R}^H=\mathbf{U}\mathbf{S}\mathbf{R}_1^H$, where $\mathbf{S}$ is a diagonal matrix containing the first $N_\mathrm{s}$ nonzero singular values, and $\mathbf{R}_1$ contains the associated singular vectors in $\mathbf{R}$. It can be shown that the equality in \eqref{inequ} is satisfied when $\mathbf{W}_{\mathrm{D},k}=\mathbf{R}_1\mathbf{U}^H$.

\subsubsection{Analog Combiners Design}
Recall the expression of the sum-MSE in \eqref{MSEequation} after the optimization of the digital precoder. Due to the constant modulus constraint on $\mathbf{W}_{\mathrm{RF},k}$ and the unitary constraint of \eqref{supposing}, the variable $\lambda$ in \eqref{MSEequation} is equal to $\lambda=\frac{\sigma^2KN_\mathrm{r}N_\mathrm{s}}{P}$. The sum-MSE in \eqref{MSEequation} under the approximation of $\mathbf{V}_\mathrm{RF}^H\mathbf{V}_\mathrm{RF}\approx N_\mathrm{t}\mathbf{I}_{N_\mathrm{t}^\mathrm{RF}}$ can be expressed as
\begin{equation}
\begin{split}
J_{\mathrm{sum}} = & \lambda\mathrm{tr} \big((\mathbf{V}_\mathrm{RF}^H\mathbf{H}^H\mathbf{W} \mathbf{W}^H \mathbf{H} \mathbf{V}_\mathrm{RF} +\lambda\mathbf{V}_\mathrm{RF}^H\mathbf{V}_\mathrm{RF})^{-1} \\ & \quad\quad\times\mathbf{V}_\mathrm{RF}^H\mathbf{V}_\mathrm{RF}\big) + KN_\mathrm{s}-N_\mathrm{t}^\mathrm{RF}\\
\approx & \eta J(\mathbf{W}_{\mathrm{RF},k}) + KN_\mathrm{s}-N_\mathrm{t}^\mathrm{RF},
\label{MSEr}
\end{split}
\end{equation}
where
\begin{equation}
\begin{split}
&J(\mathbf{W}_{\mathrm{RF},k})\\=&\mathrm{tr}\big((\mathbf{V}_\mathrm{RF}^H\mathbf{H}^H\mathbf{W}\mathbf{W}^H \mathbf{H}\mathbf{V}_\mathrm{RF}+\eta\mathbf{I}_{N_\mathrm{t}^\mathrm{RF}})^{-1}\big)\\
=&\mathrm{tr}\Big(\big(\sum\limits_{k=1}^K\overline{\mathbf{H}}_k^H\mathbf{W}_{\mathrm{RF},k} \mathbf{W}_{\mathrm{D},k}\mathbf{W}_{\mathrm{D},k}^H\mathbf{W}_{\mathrm{RF},k}^H \overline{\mathbf{H}}_k+\eta \mathbf{I}_{N_\mathrm{t}^\mathrm{RF}}\big)^{-1}\Big),
\end{split}\nonumber
\end{equation}
with $\overline{\mathbf{H}}_k=\mathbf{H}_k\mathbf{V}_\mathrm{RF}$. It turns out that it is still difficult to minimize $J\left(\mathbf{W}_{\mathrm{RF},k}\right)$ and further mathematical manipulation is needed. Thus, we introduce the following proposition.
\begin{proposition} Define $\mathbf{\Omega}=\left[
          \begin{array}{cc}
            \mathbf{I}_{N_\mathrm{s}}&\mathbf{0}\\ \mathbf{0}&\mathbf{0}\\
          \end{array}
        \right]$. It can be shown that
$\mathrm{tr}\left((\mathbf{A}^H\mathbf{{\Omega}}\mathbf{A}+\mathbf{I}_{N_\mathrm{t}^\mathrm{RF}})^{-1}\right)\geq\mathrm{tr} \left((\mathbf{A}^H\mathbf{A}+\mathbf{I}_{N_\mathrm{t}^\mathrm{RF}})^{-1}\right)$, where $\mathbf{A}$ is an $N_\mathrm{r}^\mathrm{RF}\times N_\mathrm{t}^\mathrm{RF}$ arbitrary matrix.
\label{theo1}
\end{proposition}

$\mathnormal{Proof}$:
First define two matrices $\mathbf{A}_1=\mathbf{A}^H\mathbf{{\Omega}}\mathbf{A}$ and $\mathbf{A}_2=\mathbf{A}^H(\mathbf{I}_{N_\mathrm{r}^\mathrm{RF}}-\mathbf{{\Omega}})\mathbf{A}$. It can be shown that $\mathrm{tr}\left((\mathbf{A}^H\mathbf{A}+\mathbf{I}_{N_\mathrm{t}^\mathrm{RF}})^{-1}\right)=\mathrm{tr} \left((\mathbf{A}_1+\mathbf{A}_2+\mathbf{I}_{N_\mathrm{t}^\mathrm{RF}})^{-1}\right)$. Denote the eigenvalues of $\mathbf{A}_1+\mathbf{I}_{N_\mathrm{t}^\mathrm{RF}}$ and those of $\mathbf{A}_1+\mathbf{A}_2+\mathbf{I}_{N_\mathrm{t}^\mathrm{RF}}$ by $\mu_1\leq\mu_2\ldots\leq\mu_{N_\mathrm{t}^\mathrm{RF}}$ and $\upsilon_1\leq\upsilon_2\ldots\leq\upsilon_{N_\mathrm{t}^\mathrm{RF}}$, respectively. According to the Weyl Theorem \cite{horn1990matrix}, we have $\mu_j\leq\upsilon_j$, for $j=1,\dots,N_\mathrm{t}^\mathrm{RF}$ and
\begin{equation}
\mathrm{tr}((\mathbf{A}^H\mathbf{A}+\mathbf{I}_{N_\mathrm{t}^\mathrm{RF}})^{-1})=\sum_j\frac{1}{\upsilon_j} \leq \sum_j\frac{1}{\mu_j}=\mathrm{tr}((\mathbf{A}_1+\mathbf{I}_{N_\mathrm{t}^\mathrm{RF}})^{-1}),\nonumber
\end{equation}
where the equality holds when $N_\mathrm{r}^\mathrm{RF}=N_\mathrm{s}$. \hfill $\blacksquare$

\begin{figure*}[!t]
\centering
\begin{center}
\centerline{\subfigure[]{\includegraphics[width=3.6cm]{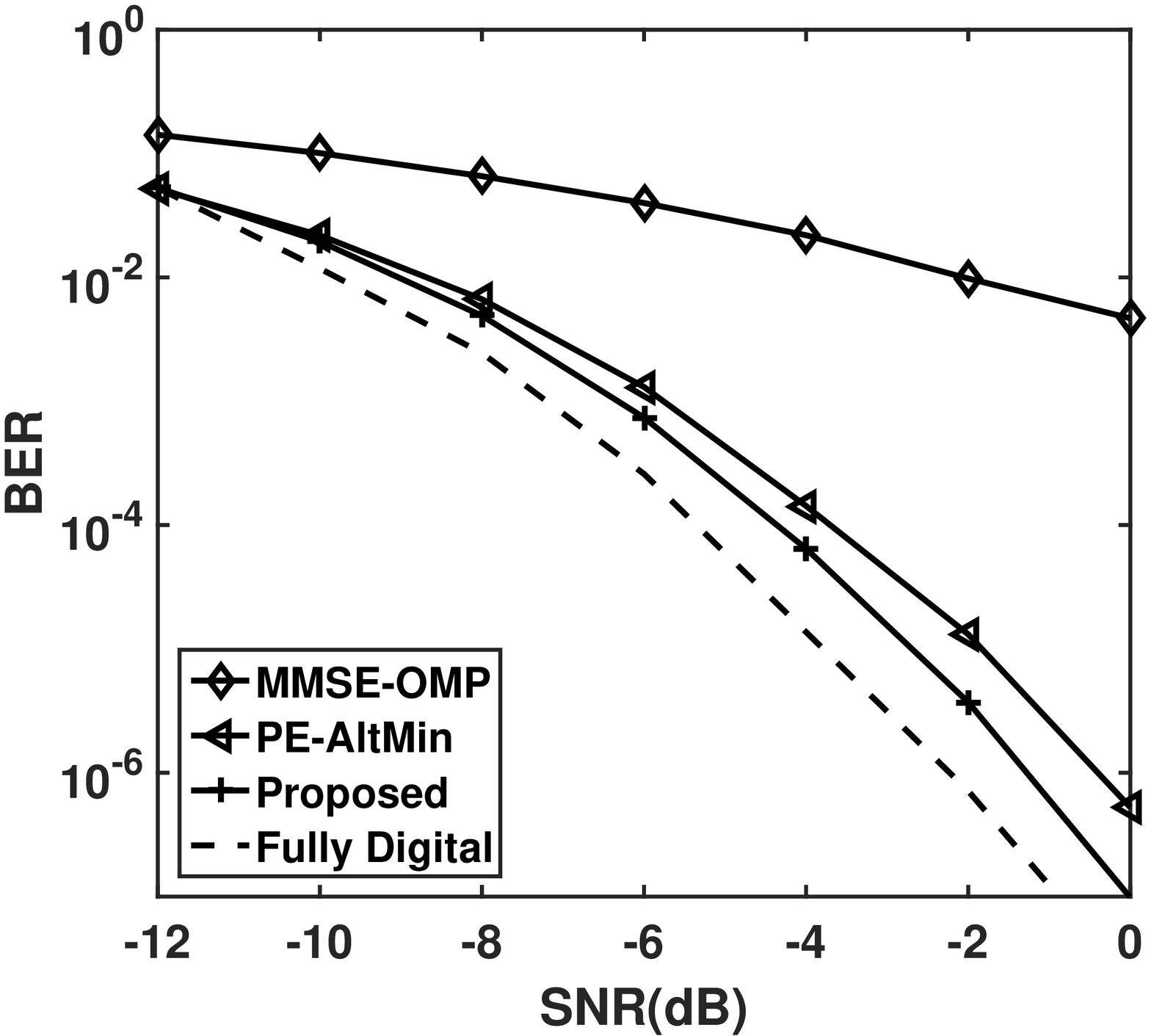}\label{fig:BERvsSNR_256_16_2_8}} \hfil \subfigure[]{\includegraphics[width=3.6cm]{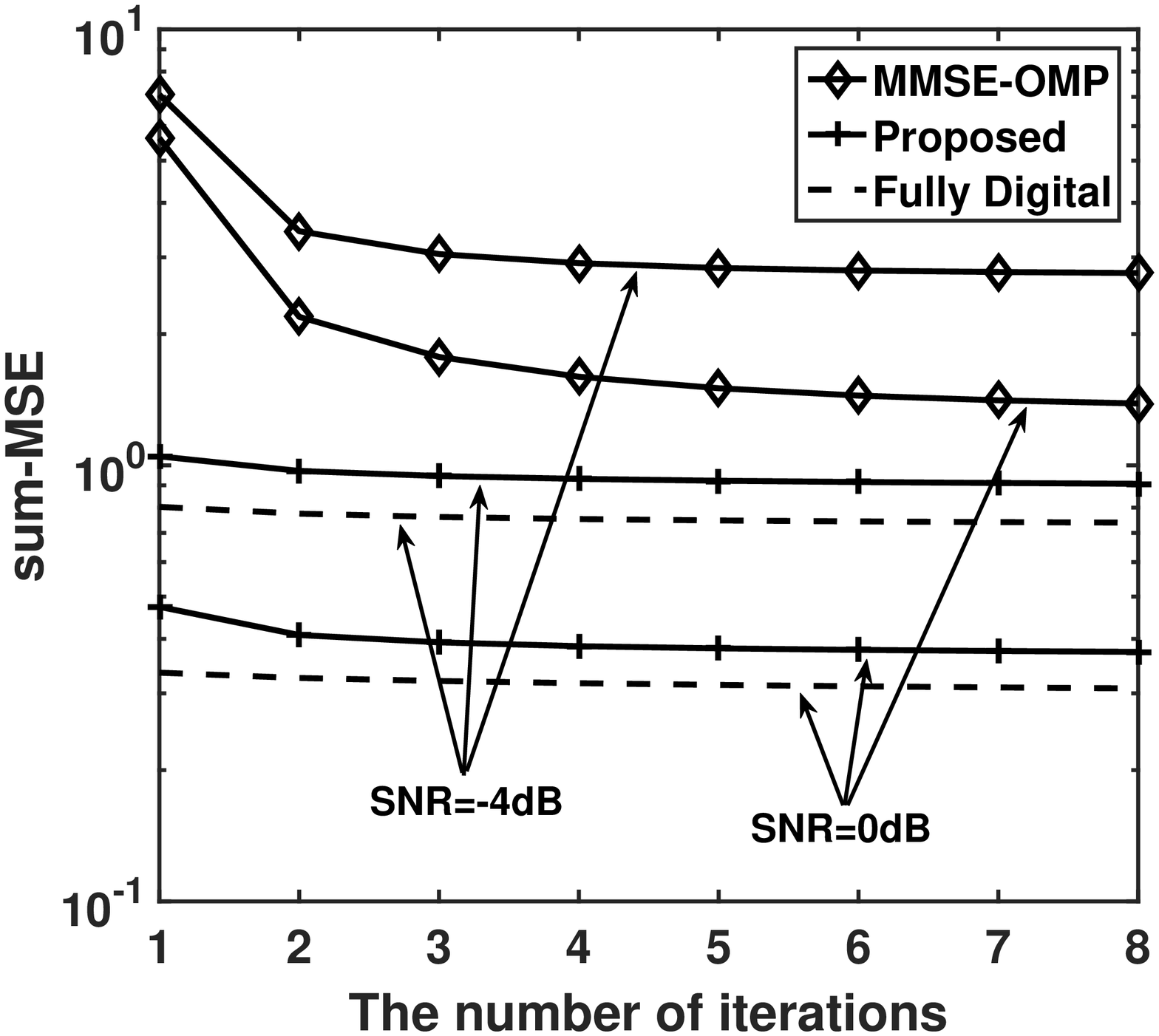}\label{fig:MSEvsNUM256_16_2_8}}\hfil \subfigure[]{\includegraphics[width=3.6cm]{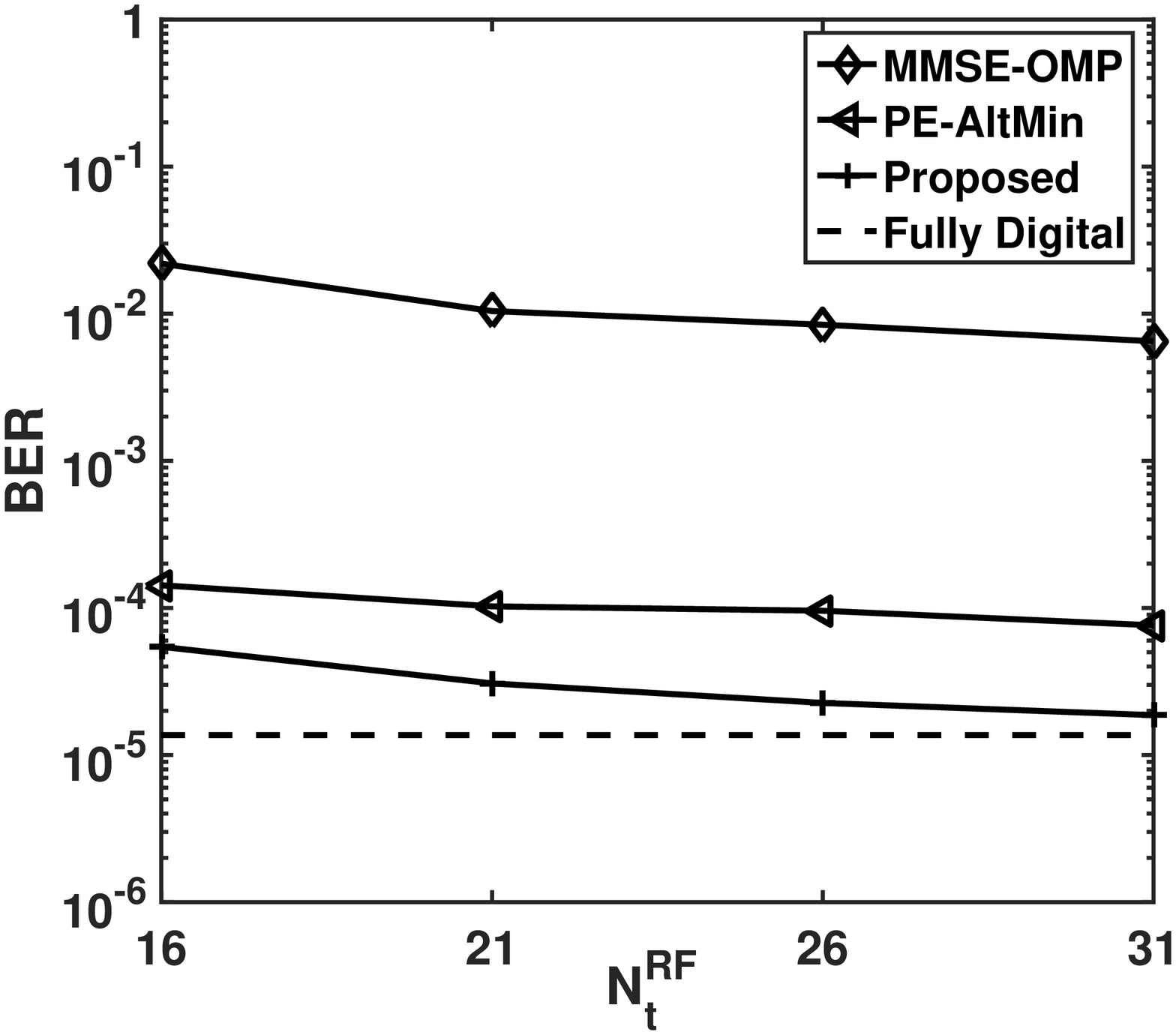}\label{fig:BERvsRFnew}}\hfil}\caption{Comparison of different beamforming schemes in an 8-user mmWave MIMO system. (a) BER v.s. SNR. (b) Sum-MSE v.s. $N_\mathrm{it}$. (c) BER v.s. $N_\mathrm{t}^\mathrm{RF}$.} \label{fig:sim}
\end{center}
\end{figure*}

By using Proposition \ref{theo1} and the fact that $\mathbf{W}_{\mathrm{D},k}\mathbf{W}_{\mathrm{D},k}^H=\mathbf{Z}_k\mathbf{\Omega}\mathbf{Z}_k^H$ from the orthogonal constraint in (\ref{supposing}), we have
\begin{equation}
\begin{split}
J(\mathbf{W}_{\mathrm{RF},k})
=&\mathrm{tr}\Big((\sum\limits_{k=1}^K\overline{\mathbf{H}}_k^H\mathbf{W}_{\mathrm{RF},k}\mathbf{Z}_k \mathbf{{\Omega}}\mathbf{Z}_k^H\mathbf{W}_{\mathrm{RF},k}^H \overline{\mathbf{H}}_k+\eta\mathbf{I}_{N_\mathrm{t}^\mathrm{RF}})^{-1}\Big)\\
\geq&\mathrm{tr}\Big((\sum\limits_{k=1}^K\overline{\mathbf{H}}_k^H\mathbf{W}_{\mathrm{RF},k} \mathbf{W}_{\mathrm{RF},k}^H\overline{\mathbf{H}}_k+\eta\mathbf{I}_{N_\mathrm{t}^\mathrm{RF}})^{-1}\Big).
\end{split}\nonumber
\end{equation}
Now the users' analog combiners can be optimized by minimizing the lower bound of $J\left(\mathbf{W}_{\mathrm{RF},k}\right)$, which is denoted by $J_\mathrm{LB}\left(\mathbf{W}_{\mathrm{RF},k}\right)$. It turns out that $\mathbf{W}_{\mathrm{RF},k}$ can be optimized column-by-column via the GEVD method. Specifically, with the definition of $\mathbf{A}_{\mathrm{r},j,k}=\sum_{f=1,f\neq k}^K\overline{\mathbf{H}}_f^H\mathbf{W}_{\mathrm{RF},f}\mathbf{W}_{\mathrm{RF},f}^H \overline{\mathbf{H}}_f+\overline{\mathbf{H}}_k^H\overline{\mathbf{W}}_{\mathrm{RF},k}^{(j)} (\overline{\mathbf{W}}_{\mathrm{RF},k}^{(j)})^H\overline{\mathbf{H}}_k+\eta\mathbf{I}_{N_\mathrm{t}^\mathrm{RF}}$, $J_\mathrm{LB}\left(\mathbf{W}_{\mathrm{RF},k}\right)$ becomes
\begin{equation}
\begin{split}
&J_\mathrm{LB}(\mathbf{W}_{\mathrm{RF},k})\\
=&\mathrm{tr}(\mathbf{A}_{\mathrm{r},j,k}^{-1})-\frac{\mathbf{w}_{\mathrm{RF},k}^{(j)H} (\overline{\mathbf{H}}_k\mathbf{A}_{\mathrm{r},j,k}^{-2}\overline{\mathbf{H}}_k^H) \mathbf{w}_{\mathrm{RF},k}^{(j)}}{\mathbf{w}_{\mathrm{RF},k}^{(j)H} (\frac{1}{N_\mathrm{r}}\mathbf{I}+\overline{\mathbf{H}}_k\mathbf{A}_{\mathrm{r},j,k}^{-1} \overline{\mathbf{H}}_k^H)\mathbf{w}_{\mathrm{RF},k}^{(j)}}.
\end{split}\nonumber
\end{equation}
By comparing it with \eqref{process_t}, it can be found that they have the same form and thus $\mathbf{w}_{\mathrm{RF},k}^{(j)}$ can be optimized in the same way. Finally, by using the alternating minimization method, the hybrid precoder and the hybrid combiners are alternatively optimized until a stop condition is satisfied.

\section{Simulation Results and Conclusion}\label{result}
Consider a multiuser ($K=8$) mmWave MIMO system with $N_\mathrm{t}^\mathrm{RF}=16$, $N_\mathrm{t}=256$, $N_\mathrm{s}=2$, $N_\mathrm{r}^\mathrm{RF}=2$ and $N_\mathrm{r}=16$. The channels are generated according to the geometric channel model in \eqref{channel} with $L=20$, $\alpha_{l,k}\sim\mathcal{CN}\left(0,1\right)$, $d=\lambda_c/2$ and uniformly randomly distributed AoAs and AoDs in $[0,2\pi]$.

Fig. \ref{fig:BERvsSNR_256_16_2_8} shows bit error rate (BER) v.s. signal to noise ratio (SNR) for the proposed HBF, the conventional phase extraction alternating minimization (PE-AltMin) HBF \cite{yu2016alternating}, the MMSE-OMP HBF \cite{nguyen2017hybrid}, and the fully digital beamforming (FDBF) schemes \cite{joung2007regularized} with quadrature phase-shift keying (QPSK) modulation. Note that both the proposed HBF scheme and the conventional MMSE-OMP and FDBF schemes apply the alternating minimization method to alternatively optimize the BS's precoder and the users' combiners. In these schemes, the iteration is stopped when the difference between the sum-MSE values in two continuous iterations is less than $10^{-6}$. For the PE-AltMin scheme, as the original HBF problem is decoupled into two matrix approximation sub-problems at the BS and users' sides \cite{yu2016alternating}, the matrices to be approximated are set to the ones in the FDBF scheme \cite{joung2007regularized}. Fig. \ref{fig:BERvsSNR_256_16_2_8} shows that the proposed HBF scheme significantly outperforms the conventional ones. This is because in the MMSE-OMP scheme the analog beamformers are limited to a predefined set consisting of only the antenna array response vectors and in the PE-AltMin scheme the original sum-MSE optimization problem is indirectly solved as it is converted into the matrix approximation problem.




Fig. \ref{fig:MSEvsNUM256_16_2_8} shows the averaged sum-MSE over 1000 channel realizations as a function of the number of iterations, $N_\mathrm{it}$, in the alternating minimization between the BS and users' sides for different schemes when $\mathrm{SNR}=-4\mathrm{dB}$ and $0\mathrm{dB}$. It can be seen that the proposed HBF scheme converges quickly to a lower sum-MSE value than the MMSE-OMP scheme, and such value is close to that of the fully digital scheme.

Fig. \ref{fig:BERvsRFnew} shows the BER performance as a function of $N_\mathrm{t}^\mathrm{RF}$ when $\mathrm{SNR}=-4\mathrm{dB}$. Here other system parameters are the same as those in Fig. \ref{fig:BERvsSNR_256_16_2_8}. It can be seen that with more RF chains the proposed HBF scheme approaches the fully digital one more quickly than other HBF schemes.

Comparing the computational complexity of different schemes in terms of the number of complex multiplications, the complexity of MMSE-OMP is in the order of $\mathcal{O}(N_\mathrm{it}(N_\mathrm{t}^\mathrm{RF}N_\mathrm{t}^3+KN_\mathrm{t}^3))$, as shown in \cite{nguyen2017hybrid}, and that of FDBF is $\mathcal{O}(N_\mathrm{it}(N_\mathrm{t}^3+KN_\mathrm{r}^3))$ because of the matrix inversion at both the BS and the users. The PE-AltMin scheme needs at least the complexity of FDBF to obtain the target fully digital matrices. The complexity of the proposed scheme is mainly in GEVD, which is in the order of $\mathcal{O}(N_\mathrm{it}(N_\mathrm{t}^{\mathrm{RF}}N_\mathrm{t}^3+KN_\mathrm{r}^{\mathrm{RF}}N_\mathrm{r}^3))$. However, it can be reduced to $\mathcal{O}(N_\mathrm{it}(N_\mathrm{t}^{\mathrm{RF}}N_\mathrm{t}^2+KN_\mathrm{r}^{\mathrm{RF}}N_\mathrm{r}^2))$  by using the power method \cite{horn1990matrix} since only the largest generalized eigenvector needs to be computed. Thus, the complexity of the proposed HBF scheme is not more than that of the conventional HBF schemes.

In conclusion, we have proposed an MMSE HBF scheme for multiuser MIMO mmWave systems based on the alternating minimization method. In particular, we showed that the RF beamformers can be optimized via GEVD. Simulation results showed that the proposed HBF scheme is able to approach the performance of the fully digital beamforming scheme.





\end{document}